\documentclass{jfm}
\usepackage{amsmath,bm,amsfonts,amssymb,upmath}
\usepackage{graphicx}
\title{Nonlinear elastic polymers in random flow}
\author[M. Martins Afonso and D. Vincenzi]{M.\ns M\ls A\ls R\ls T\ls I\ls N\ls S\ns A\ls F\ls O\ls N\ls S\ls O$^1$\ns
\and D.\ns V\ls I\ls N\ls C\ls E\ls N\ls Z\ls I$^{2,3,4}$}
\affiliation{$^1$INFM-Dipartimento di Fisica, Universit\`a di Genova
and INFN, Sezione di Genova,
Via~Dodecaneso 33, 16146 Genova, Italy
\\[\affilskip]
$^2$Dipartimento di Fisica, Universit\`a di Genova,
Via Dodecaneso 33, 16146 Genova, Italy
\\[\affilskip]
$^3$Max-Planck-Institut f\"ur Dynamik und
Selbstorganisation, G\"ottingen, Germany
\\[\affilskip]
$^4$LASSP, Clark Hall, Cornell University, Ithaca, NY 14853, USA
}
\date{\today}

\usepackage{natbib}
\allowdisplaybreaks[1]
\begin{document}

\maketitle

\begin{abstract}
Polymer stretching in random smooth flows is investigated
within the framework of the FENE dumbbell model.
The advecting flow is Gaussian and short-correlated
in time. 
The stationary probability density function of polymer
extension is derived exactly. 
The characteristic time needed for the system to attain the 
stationary regime is computed as a function of the Weissenberg number 
and the maximum length of polymers.
The transient relaxation to the stationary regime is predicted to be
exceptionally slow in the proximity of the coil--stretch transition.
\end{abstract}

\section{Introduction}

The ability of polymers to considerably change the large-scale
statistics of the advecting flow has  important practical
applications, drag reduction being one of the most relevant ones
\cite[see][]{Gyr}. Polymers affect the dynamics of the
advecting velocity field only if they are highly elongated.
Understanding how a single polymer chain is stretched by a
random flow is thus the first issue to address in the study of
hydrodynamical properties of polymer solutions.

At equilibrium, the radial shape of coiled polymers is spherical due to
their entropy.
When placed in a non-homogeneous flow, polymers
are deformed and stretched by the gradients of the velocity. The
product of the longest relaxation time of polymers and the
characteristic rate of deformation is called the Weissenberg
number $\textit{Wi}$. For small~$\textit{Wi}$ the entropic force prevails and polymers
are in the coiled state. When $\textit{Wi}$ exceeds a critical value the
molecules become highly elongated and their extension increases
sharply. This phenomenon is called coil--stretch transition and the
critical Weissenberg number is known to be approximately one.

We investigate the statistics of polymer extension in the FENE 
(finite extensible nonlinear elastic)
dumbbell model
 \cite[see e.g.][for a review]{Bird}. 
A polymer 
is described as two beads joined by an elastic spring.
The elastic force diverges as the elongation
of polymers attains its maximum value, $R_m$, and this gives a large-extension cutoff.
Consequently the stationary regime exists however strong the velocity gradients are.
Since attention is directed only to
the dynamics of a single molecule,
the feedback on the advecting flow is
disregarded. To allow analytical progress, 
the random flow is chosen to
have the Batchelor-Kraichnan statistics.
This means that the considered flow is Gaussian, white in time and linear in space.
The Batchelor-Kraichnan model is a fully solvable
model for passive turbulent transport which
can provide useful connections between theory and real behaviours
\cite*[see][for a review on the applications
to scalar and magnetic fields]{Falkovich}. 
The results should be intended as a qualitative description of real polymer dynamics.
We derive the complete form of the stationary probability density function (p.d.f.)
of polymer extension and describe how the statistics of polymer stretching
changes with increasing velocity gradients.
Concerning the statistics at finite times, 
we compute the typical time needed for the system to reach
the steady state and predict how it depends on the 
maximum length of polymers and the stretching by the flow.
Our analysis shows that 
the coil--stretch transition of polymers is characterised 
by an anomalous dynamics in time.

The coil--stretch transition was predicted in 1974
for shear and hyperbolic flows \cite[][]{DeGennes} 
and has been widely
studied experimentally for such flows  
\cite*[][]{Perkins,Smith,Hur,Babcock}. In contrast, the experimental
study of polymer dynamics in random flows is a very recent
achievement. This is due to the difficulty in generating a flow
that is random at scales comparable with the size of polymers
(about 100 $\umu$m). This difficulty can be overcome thanks to the
elastic turbulence discovered by \cite{Groisman}; the flow of a
highly elastic polymer solution at low Reynolds numbers, but large
$\textit{Wi}$ has all the main properties of fully developed turbulence.
Therefore, in solutions of sufficiently elastic polymers it is
possible to excite turbulent motion in exceedingly small volumes.
Exploiting elastic turbulence in polymer solutions,
\cite*{Gerashchenko} thus investigated the stretching and the
deformation of a single DNA molecule in a three-dimensional random flow. 

Theoretical studies concerning the coil--stretch transition
in random flows focused mainly on the Hookean dumbbell model
\cite*[][]{Chertkov,Balkovsky,Celani}.
This model is suitable only for the coiled state ($\textit{Wi}<1$)
since
the linear force in principle allows infinite extensions and 
for large~\textit{Wi} polymers can become more
and more elongated under the action of velocity gradients.
For $\textit{Wi}\ge 1$ a stationary p.d.f. of the extension
no longer exists and
this behaviour was conjectured to coincide with the coil--stretch transition.
To overcome this oversimplification,
the maximum length of polymers must be
taken into account. One possibility is  to replace the Hookean force by 
a nonlinear elastic force.
\cite{Chertkov} obtained 
the large-value tail of the stationary p.d.f. of the extension
for a general anharmonic force; such approximate analysis was subsequently
applied
by \cite{Thiffeault} to the FENE model.
Here we exactly derive the complete statistics of polymer stretching
within the context of the FENE dumbbell model at general \textit{Wi}.

Concerning the statistics at finite times,
preliminary results were obtained for the Hookean model 
by \cite{Celani}. There,
the relaxation time to the stationary regime could be defined
only in the coiled state. \cite{Celani}
thus derived the behaviour of the 
relaxation time for very
small~\textit{Wi} and observed a divergence for $\textit{Wi}=1$;
this suggested a critical behaviour close to the coil--stretch
transition. 
We present the first prediction of the complete dependence of
the transient relaxation time on~$\textit{Wi}$ and~$R_m$
with the more realistic FENE model.

The rest of this paper is organised as follows. In
\S\ref{sec:main} we introduce the model and present the main
results. The stationary p.d.f. of the elongation and the 
transient relaxation
time are computed in \S\ref{sec:staz} and in \S\ref{sec:relax}
respectively. In \S\ref{sec:conclusions} we discuss the relevance of our 
results for experiments and numerical simulations.

\section{Coil--stretch transition}
\label{sec:main}

In elastic dumbbell models a polymer is described as two beads connected by a spring.
The beads represent the ends of the molecule and their separation is a measure of the extension.
The beads experience: (a) a hydrodynamic drag force modeled by the Stokes law;
(b) a Brownian force due to thermal fluctuations of the fluid; (c)
an elastic force due to the spring connecting one bead to the other.
We consider two-dimensional and three-dimensional flows indifferently,
the dimension of the flow being denoted by~$d$.
Since in physical applications the elongation of polymers
is always smaller than the viscous scale
of the flow, the dumbbell is assumed to
move in a linear velocity field
$\bm v(\bm r,t)=\bm v_0(t) +\bm r\cdot\bm\nabla \bm v(t)$.
Inertial effects and hydrodynamic interactions between the beads are neglected.
Consequently, the separation vector between the beads, $\bm R$,
evolves according to the stochastic differential equation \cite[see][]{Bird}
\begin{equation}
\label{eq:dumbbell}
d\bm R=(\bm R\cdot\bm\nabla\bm v)dt-
\frac{\bm F(\bm R)}{\tau}\, dt+\sqrt{\frac{2R_0^2}{\tau}}\,d\bm W,
\end{equation}
where $R_0$ is the equilibrium length of the polymer, $\tau$ is
its relaxation time in the absence of flow, and $\bm W$ is a
$d$-dimensional Brownian motion which accounts for thermal noise.
In the FENE dumbbell model,
the elastic force $\bm F(\bm R)$ takes the form
$\bm F(\bm R)=\bm R/(1-R^2/R_m^2)$, where $R_m$ denotes the maximum extension
of the molecule. 
In physical applications the ratio
$R_m/R_0$ usually lies between~10 and~100 \cite[][]{Bird}. 
The length of the vector $\bm R$ is a measure of the extension of the
polymer.

Within the Kraichnan model
$\bm v(\bm x,t)$ is a statistically
homogeneous Gaussian field with zero mean and
second-order correlation \cite[][]{Kraichnan_68}
\[
\langle
v_i(\bm x,t)\, v_j(\bm x +\bm r,t')\rangle=
{\mathsfbi D}_{ij}(\bm r)\,\delta(t-t').
\]
In the so-called Batchelor regime the flow is assumed to be smooth in space.
If we further impose incompressibility and
statistical invariance with respect to reflections and rotations,
the tensor ${\mathsfbi D}_{ij}(\bm r)$ must take the form \cite[][]{Yaglom}
\[
{\mathsfbi D}_{ij}(\bm r)=D_0\delta_{ij}-D_1[(d+1)\delta_{ij}r^2-2r_ir_j],
\]
where $D_0$ represents the eddy diffusivity of the flow and $D_1$
determines the intensity of turbulent fluctuations. In random
flows the Weissenberg number can be defined as
$\textit{Wi}=\lambda\tau$, where $\lambda$ is the maximum Lyapunov
exponent of the flow, that is the average logarithmic growth rate
of nearby fluid particle separations. The maximum Lyapunov
exponent of the Batchelor-Kraichnan flow has asymptotically a
Gaussian p.d.f. with mean value $\lambda=D_1d(d-1)$ and variance
$\Delta=2\lambda/d$ \cite[][]{Kraichnan_74}. 

\subsection{Stationary regime}

The statistics of polymer elongation is described by the p.d.f. of
the norm of $\bm R$ averaged over velocity
realizations\footnote{Because of the statistical homogeneity of
$\bm v$, the average p.d.f. of the elongation does not depend on the
point of application of the vector $\bm R$.}: ${\cal
P}(R,t)=\int\langle P(\bm R,t)\rangle R^{d-1}\, d\Omega$, where
$d\Omega$ denotes integration over angular variables. When the
flow $\bm v$ has the Batchelor-Kraichnan statistics ${\cal
P}(R,t)$ obeys a one-dimensional Fokker-Planck equation with
nontrivial drift and diffusion coefficients (see
\S\ref{sec:staz}). Under reflecting boundary conditions (the
probability does not flow outside the domain of definition) the
system reaches a steady state for all \textit{Wi}\footnote{This should be
contrasted with the Hookean model where the stationary regime does
not exist for $\textit{Wi}\ge 1$
\cite[see][]{Chertkov,Balkovsky,Thiffeault,Celani}.}. The stationary p.d.f.
of the elongation, ${\cal P}_{\mathrm{st}}(R)=
\lim_{t\to\infty}{\cal P}(R,t)$, has the form (see
\S\ref{sec:staz})
\begin{equation}
\label{eq:staz}
{\cal P}_{\mathrm{st}}(R)=N\,
R^{d-1}\Big(1+\frac{\textit{Wi}}{d}\frac{R^2}{R_0^2}\Big)^{-h}
\Big(1-\frac{R^2}{R_m^2}\Big)^h
\qquad 0\le R\le R_m
\end{equation}
where $h=[2(R_0^2/R_m^2+\textit{Wi}/d)]^{-1}$ and $N$ is the
normalization coefficient (see~\eqref{eq:norm} below). The
stationary p.d.f. is shown in figure~\ref{fig:staz} for different $\textit{Wi}$.
For small elongations compared to the equilibrium length, ${\cal
P}_{\mathrm{st}}(R)$ scales as $R^{d-1}$; this result holds for a general
elastic force since the left tail of ${\cal P}_{\mathrm{st}}$ comes
from the events where the elastic force dominates and
equation~\eqref{eq:dumbbell} reduces to a $d$-dimensional Langevin
equation (for physically meaningful elastic interactions $\bm
F(\bm R)$ should scale as $\bm R$ for $R\to 0$). For intermediate extensions,
$R_0\ll R\ll R_m$, the stationary p.d.f. is proportional to the power law 
$R^{d-1-2h}$ in accordance with the prediction of \cite{Balkovsky}.
For large
elongations  ${\cal
P}_{\mathrm{st}}(R)$ scales as $(R_m^2-R^2)^h$
and vanishes for $R=R_m$. In practical applications, $R_0/R_m\ll
1$, the exponent $h$ is approximatively $d/(2 \textit{Wi})$, 
as predicted by \cite{Thiffeault}. Obviously,
when $R_m\to\infty$ and $\textit{Wi}<1$, $\mathcal{P}_{\mathrm{st}}(R)$  tends to the
stationary solution of the Hookean model
\cite[see][]{Celani}.

\begin{figure}
\centering
\includegraphics[height=5cm]{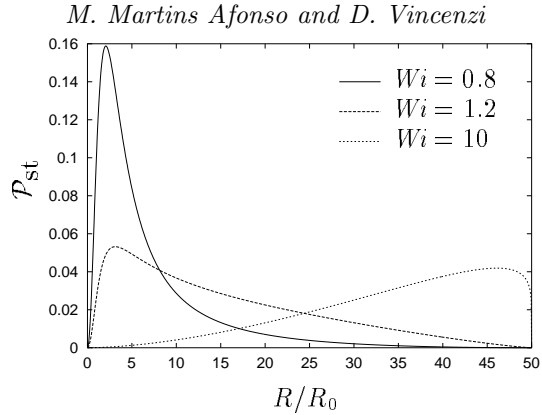}
\caption{ Stationary p.d.f. of polymer elongation for the
three-dimensional FENE  model at different Weissenberg numbers
$\textit{Wi}$ ($R_m/R_0=50$).} \label{fig:staz}
\end{figure}

\begin{figure}
\centering
\hspace*{0.55cm}\includegraphics[height=4cm]{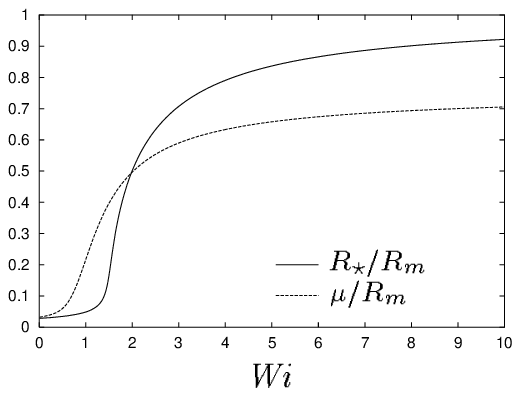}\hfill
\includegraphics[height=4cm]{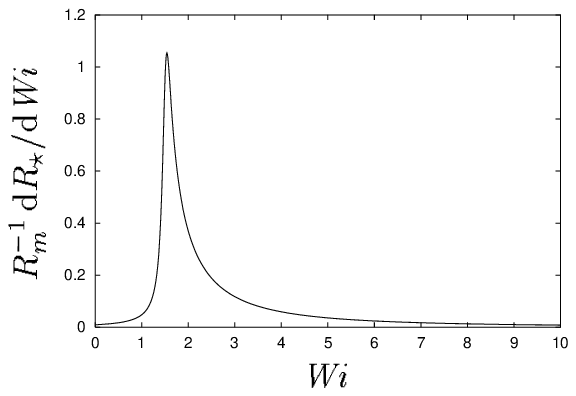}
\caption{ Left: Most probable rescaled elongation $R_\star/R_m$
and average rescaled extension $\mu/R_m$ as functions of the
Weissenberg number $\textit{Wi}$ ($d=3$, $R_m=50$, $R_0=1$). Right:
First derivative of $R_\star/R_m$ with respect to $\textit{Wi}$.}
\label{fig:max}
\end{figure}

The maximum of the p.d.f., $R_\star$, determines the fraction of
polymers which are highly stretched. The graph of $R_\star$ as a
function of $\textit{Wi}$ is shown in figure~\ref{fig:max}. When $\textit{Wi}$ is
smaller than one, $R_\star$ is of the order of $R_0$ and most of
polymers have the coiled equilibrium configuration. With
increasing $\textit{Wi}$ the most probable elongation $R_\star$ grows slowly
until~$\textit{Wi}$ exceeds $d/(d-1)$. Then, a sharp transition
occurs to a strongly elongated state. This can be appreciated from
the behaviour of the first-order derivative of $R_\star$ as a
function of~$\textit{Wi}$ (figure~\ref{fig:max}). As $\textit{Wi}$ becomes very
large, $R_\star$ approaches $R_m$. 
The same analysis holds for the
average extension $\mu$, apart from the fact that it starts increasing
for a smaller $\textit{Wi}$ and 
its limiting value is $\frac{3}{4}R_m$ (see figure~\ref{fig:max}).
It is worth noticing that the coil--stretch transition becomes
sharper and sharper with increasing $R_m$ (not plotted).

\begin{figure}
\centering
\includegraphics[height=4cm]{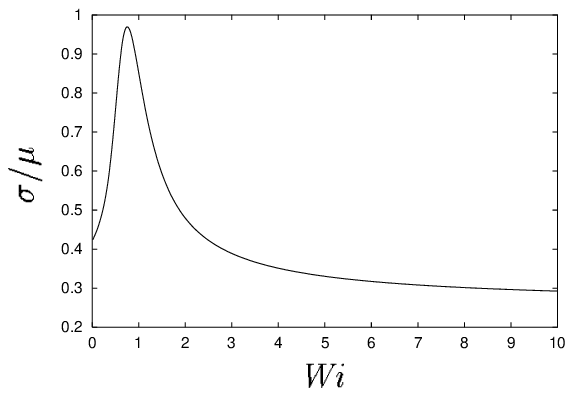}\hfill
\includegraphics[height=4cm]{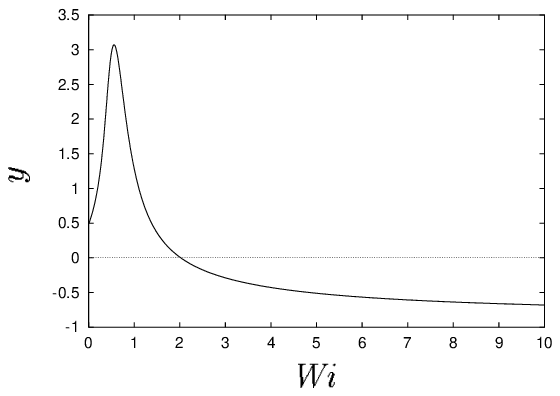}
\caption{ Left: Normalized root mean square $\sigma/\mu$ as a
function of the Weissenberg number $\textit{Wi}$ ($d=3$, $R_m=50$,
$R_0=1$). Right: Skewness $y$ \textsl{vs} $\textit{Wi}$ for the same
values of the parameters.} \label{fig:sigmaskew}
\end{figure}

The normalized r.m.s. value of the extension, $\sigma/\mu$,
$\sigma^2=\int(R-\mu)^2{\cal P}_{\mathrm{st}}(R)\,\mathrm{d}R$, 
is represented in figure~\ref{fig:sigmaskew}. It increases at
low $\textit{Wi}$ until it reaches a maximum value; then $\sigma$
is
compensated by the sharp increase in $\mu$ and
at large $\textit{Wi}$ the rescaled r.m.s.
eventually relaxes to the constant value $1/\sqrt{15}$.

The skewness
$y=[\int(R-\mu)^3{\cal P}_{\mathrm{st}}(R)\,\mathrm{d}R]/\sigma^3$ is positive for small
$\textit{Wi}$ and becomes negative at large $\textit{Wi}$ (figure~\ref{fig:sigmaskew}) 
accordingly with the qualitative behaviour of the stationary p.d.f.
(figure~\ref{fig:staz}). The maximum of skewness in the neighbourhood of the
coil--stretch transition can be easily understood as follows.
At low $\textit{Wi}$ the p.d.f. is peaked at $R_0$ and the skewness is positive. With
increasing $\textit{Wi}$ the right tail starts raising, but $\mu$
is still of the order of $R_0$: the skewness, therefore, increases
and achieves is maximum value. Beyond the coil--stretch transition 
$\mu$ starts moving towards the maximum extension and the skewness decreases
until it becomes negative at large $\textit{Wi}$, 
that is when the p.d.f. has a long left tail.
The limiting value of  $y$ for
$\textit{Wi}\to\infty$ is $-\frac{2}{3}\sqrt{5/3}$.

\subsection{Relaxation to the stationary regime}

We now turn to the time dependence of the p.d.f. of the elongation.
Starting from an initial condition peaked at $R_0$, the system relaxes to
the stationary regime described by~\eqref{eq:staz}. The time needed
to reach the stationary regime, $T$,
is solution of a transcendental equation which involves
continued fractions (see \S\ref{sec:relax}).

\begin{figure}
\centering
\includegraphics[height=4.5cm]{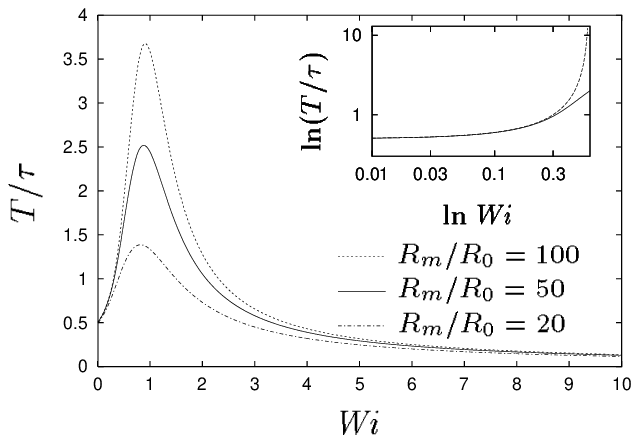}\hfill
\includegraphics[height=4.5cm]{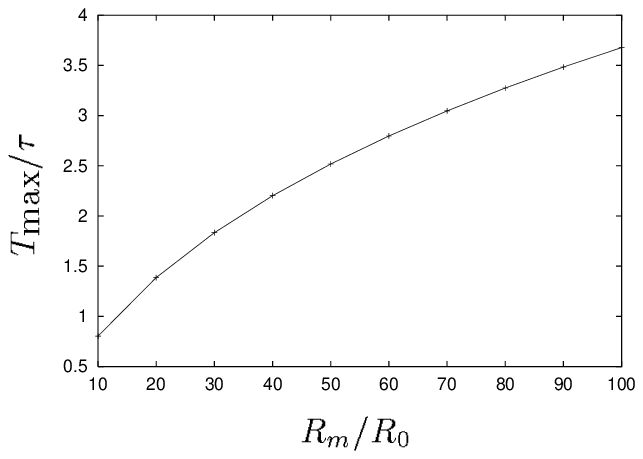}
\caption{ Left: Rescaled time of relaxation to the stationary
regime, $T/\tau$, as a function of the Weissenberg number
$\textit{Wi}$ for three different values of $R_m/R_0$ ($d=3$). The
inset shows the linear-model approximation given by
equation~\eqref{eq:lma} (dashed line) for $R_m/R_0=50$: the agreement
is good up to $d/(d+4)\simeq0.4$. Right: Dependence of the maximum
rescaled relaxation time $T_{\mathrm{max}}/\tau$ on the maximum
relative extension of polymers $R_m/R_0$.} \label{fig:Tmax}
\end{figure}

For small Weissenberg numbers, $0\le \textit{Wi}\lesssim d/(d+4)$,
the transient relaxation time
$T$ behaves according to the
prediction of the linear model \cite[][]{Celani}:
\begin{equation}
\label{eq:lma} T/\tau={\textstyle\frac{1}{2}}\left[1-\textit{Wi}\,
(d+2)/d\right]^{-1}
\end{equation}
independently of $R_m$
(see inset in figure~\ref{fig:Tmax}).

In the proximity of the coil--stretch transition $T$ displays a
maximum as a function of the $\textit{Wi}$. The relaxation is exceptionally
slow in this range of $\textit{Wi}$ because  the stationary regime results
from the competition between the coiled state and the highly
stretched state. The position and the value of the maximum
relaxation time $T_{\mathrm{max}}$ depend on the cutoff $R_m$
(figure~\ref{fig:Tmax}). As the maximum allowed extension of
polymers increases, $T_{\mathrm{max}}$ is closer and closer to
$\textit{Wi}=1$ and grows; at large $R_m$ the FENE model should
indeed match the Hookean model, where $T$ diverges as $\textit{Wi}$ tends to
one \cite[][]{Celani}.

For very large Weissenberg numbers the stretching time is small
compared to $\tau$ and the molecules are expected to rapidly reach
the highly stretched configuration. Hence, $T$ vanishes  as
$\textit{Wi}$ tends to infinity. A numerical fit shows that $T$
scales as~$\textit{Wi}^{-1}$ at large~$\textit{Wi}$.

In the next sections we explicitly derive~\eqref{eq:staz} and
the equation for $T$.

\section{Fokker-Planck equation}
\label{sec:staz}

For a fixed realization of the velocity field
the p.d.f. of
the end-to-end vector, $P(\bm R,t)$, satisfies the Fokker-Planck
equation associated with~\eqref{eq:dumbbell} \cite[see e.g.][]{Risken}:
\begin{equation}
\label{eq:FP}
\partial_t P+\mbox{div}_{\bm R}
\Big[
\Big(
\bm R\cdot\bm\nabla\bm v - \frac{\bm F (\bm R)}{\tau}\Big)P\Big]=
\frac{R_0^2}{\tau}\,
\nabla^2_{\bm R}
P.
\end{equation}
To obtain an equation for ${\cal P}(R,t)$, we have to
average  the above equation over the velocity realizations and
integrate the result over angular variables. The terms
of the type $\langle v_i P\rangle$ 
in general do not lead to a closed form for the mean p.d.f. and a closed equation cannot be deduced
from~\eqref{eq:FP}.
The Gaussianity and the $\delta$-correlation in time of the 
Batchelor-Kraichnan model provide an exact closure.
Exploiting the Novikov-Furutsu formula \cite*[see e.g.][]{Klyatskin},
we obtain: 
$
\langle (\nabla_i v_j) P\rangle=-C_{ijk\ell}\, \partial_{R_\ell}
[R_k\langle\, P\rangle]$ with 
$C_{ijk\ell}=D_1[(d+1)\delta_{ik}\delta_{j\ell}-\delta_{ij}\delta_{k\ell}-
\delta_{i\ell}\delta_{jk}]$.
We can thus derive from \eqref{eq:FP} a one-dimensional Fokker-Planck equation for ${\cal P}(R,t)$:
\begin{equation}
\label{eq:FP_average}
\partial_s {\cal P}(R,s)=-\partial_R\,
[A(R){\cal P}(R,s)]
+\partial^2_R\, [B(R){\cal P}(R,s)],
\end{equation}
where the time has been rescaled with $\tau$, $s=t/\tau$, and the
drift and diffusion coefficients have the form
\begin{equation}
\label{eq:coeff}
A(R)=\frac{(d+1)}{d}\,\textit{Wi}\, R-F(R)+(d-1)\frac{R_0^2}{R}
\qquad
\mbox{and}
\qquad
B(R)=\frac{\textit{Wi}}{d}R^2+R_0^2.
\end{equation}
The coefficients $A$ and $B$ are time independent
due to the stationarity of the advecting flow.
If $R_0$ is set to zero, then
equation~\eqref{eq:FP_average} reduces to the approximate equation
for the large-value tail of the p.d.f. derived by
\cite{Chertkov}. 

To solve~\eqref{eq:FP_average}
we impose reflecting boundary conditions, that is
the probability current associated with the solution, $J(R,s)=
A(R){\cal P}(R,s)-\partial_R\,[B(R){\cal P}(R,s)]$,
vanishes in $R=0$ and $R=R_m$ for all $s\ge 0$.
This means that there is no flow of probability through the boundaries of the
domain.
Under these conditions the stationary p.d.f. of the elongation
takes the form \cite[][]{Risken}
\begin{equation}
\label{eq:formula_staz}
{\cal P}_{\mathrm{st}}(R)=
\frac{C}{B(R)}\,\exp\bigg[\int^R_{R_1} A(x)/B(x)\,dx\bigg]
\end{equation}
where the constant $C$ and the lower integration limit $R_1$ are fixed
by the normalization condition.
The above formula holds for a general elastic force of the form
$\bm F(\bm R)=f(R)\bm R$.
Replacing the force of the FENE model
into~\eqref{eq:formula_staz}, we thus obtain~\eqref{eq:staz} with
\begin{equation}
\label{eq:norm}
N=\frac{2\,\Gamma\big(d/2+h+1\big)}
{R_m^d\,\Gamma\big(d/2\big)\Gamma(h+1)\,
_2F_1\big(d/2,h;d/2+h+1;
- \textit{Wi}\, R_m^2/ d R_0^2\big)}.
\end{equation}
The function $_2F_1$ in the normalization coefficient denotes the
hypergeometric function.

\section{Relaxation time}
\label{sec:relax}
The time dependent solution of the Fokker-Planck~\eqref{eq:FP_average}
can be obtained by separation of variables \cite[][]{Risken}. In other words,
${\cal P}(R,s)$ can be sought in the form
\begin{equation}
\label{eq:series}
{\cal P}(R,s)={\cal P}_{\mathrm{st}}(R)+
\sum_{k=1}^{\infty}c_k e^{-\mu_k s}p_k(R)
\end{equation}
where the coefficients $c_k$ are fixed by the initial condition
${\cal P}(R,0)$ and $\mu_k$, $p_k(R)$ are respectively the
eigenvalues and the eigenfunctions of the ordinary differential
equation
\begin{equation}
\label{eq:autov}
\frac{d^2}{dR^2}[B(R)\, p_k(R)]-
\frac{d}{dR}[A(R)\, p_k(R)]
+\mu_k \, p_k(R)=0.
\end{equation}
The above equation should be solved 
with reflecting boundary conditions: $J_k(0)=\lim_{R\to R_m}J_k(R)=0$, $J_k$
being the probability current associated with the eigenfunction~$p_k$. It can be shown that 
the eigenvalues $\mu_k$ are real and non-negative,
$\mathcal{P}_{\mathrm{st}}(R)$ belonging to the eigenvalue~$\mu_0=0$
\cite[][]{Risken}. As we will see, the
eigenvalues form a countable set and may be arranged in ascending
order: $0<\mu_1<\mu_2<\dots$. The reciprocal of $\mu_1$,
therefore, is the time of relaxation to the stationary regime
rescaled by~$\tau$.

Equation~\eqref{eq:autov}
is a second-order linear differential
equation with four regular singularities in the complex plane.
By the change of dependent and
independent variable
$z=(R/R_m)^2$, $p_k(z)=z^{(d-1)/2}\,(1-z)^h \, w_k(z)$,
such equation can be transformed into a standard Heun equation
for the function $w_k(z)$ \cite[see][for a review]{Ronveaux}:
\begin{equation}
\label{eq:Heun}
\frac{d^2 w_k}{dz^2}+
\left(\frac{\gamma}{z}+\frac{\delta}{z-1}+\frac{\epsilon}{z-a}
\right)\frac{dw_k}{dz}+
\frac{\alpha\beta z-q}{z(z-1)(z-a)}\, w_k=0,
\end{equation}
where
\begin{eqnarray*}
a=-\dfrac{d}{\textit{Wi}}\,\dfrac{R_0^2}{R_m^2}&&
q=\frac{d}{2}\left(h+\frac{\mu_k}{2\textit{Wi}}\right)
\\[0.3cm]
\alpha= h+\frac{d}{4}-\frac{1}{4}\sqrt{d\left(d-\frac{4\mu_k}{\textit{Wi}}
\right)}&&
\beta= h+\frac{d}{4}+\frac{1}{4}\sqrt{d\left(d-\frac{4\mu_k}{\textit{Wi}}
\right)}
\\[0.3cm]
\gamma=d/2  &\qquad \delta=h\qquad &\epsilon=1+h
\end{eqnarray*}
with $h=[2(R_0^2/R_m^2+\textit{Wi}/d)]^{-1}$. 
Reflecting boundary
conditions for $p_k$ map into the following  limiting conditions
for $w_k$: 
\begin{equation}
\label{eq:bc_w}
\lim_{z\to 0}z^{\gamma-1}w_k(z)=0 \qquad \mbox{and} \qquad
\lim_{z\to 1}(1-z)^{\delta-1}w_k(z)=0.
\end{equation}
The Heun equation is the general Fuchsian equation with four singularities.
In the standard form~\eqref{eq:Heun} the singular points are $0,1,a,\infty$.
Let $z_0$ be a generic singularity of~\eqref{eq:Heun}.
From the theory of Fuchsian equations, the local behaviour of $w_k(z)$
near $z_0$ is specified by the characteristic exponents $\rho_1$, $\rho_2$
associated with $z_0$ \cite[see e.g.][]{WW}.
If $\rho_1-\rho_2$ is not integer,
in a neighbourhood of $z_0$ which excludes the nearest other singularity
$w_k(z)$ can be written
in the form $b_1(z-z_0)^{\rho_1}\varphi(z-z_0)+b_2
(z-z_0)^{\rho_2}\psi(z-z_0)$, where $b_1$, $b_2$ are constant and $\varphi$,
$\psi$ are analytic functions such that $\varphi(z_0)\neq 0$, $\psi(z_0)\ne 0$.
If $\rho_1-\rho_2$ is integer and $\rho_1\ge\rho_2$, 
the function $\psi$ can be no longer analytic in $z_0$
and involve the function $\log(z-z_0)$.

The singularity $z=0$ has characteristic exponents 0 and $1-\gamma$; the singularity
$z=1$ has characteristic exponents 0 and $1-\delta$. In physical applications we can 
exclude the situation where $\delta$ is integer. On the contrary, $1-\gamma$ is
zero when $d=2$.

Consider first the case $d=3$, where 
there are not logarithmic singularities in $z=0$.
To fulfill conditions~\eqref{eq:bc_w}, $w_k$
must be simultaneously a local solution
about $z=0$ and $z=1$, in both cases belonging to
the exponent 0. Such a solution is called a Heun function of class I
relative to the points 0 and 1, and  exists only for a countable set
of values of $q$ and hence of $\mu_k$ \cite[see][]{Ronveaux}.
The condition for the aforementioned Heun function to exist
leads to a transcendental equation for the eigenvalues $\mu_k$ \cite[][]{Erd}:
\begin{equation}
\label{fraz_cont}
L_0-\dfrac{M_0K_1}{L_1-\dfrac{M_1K_2}{L_2-\dots}}=0,
\end{equation}
where
\begin{eqnarray*}
K_i&=&\dfrac{(i+\alpha-1)(i+\beta-1)(i+\gamma-1)(i+\omega-1)}%
      {(2i+\omega-1)(2i+\omega-2)}\\[0.2cm]
L_i&=&q+ai(i+\omega)-\dfrac{\epsilon\, i(i+\omega)(\gamma-\delta)+
[i(i+\omega)+\alpha\beta][2i(i+\omega)+\gamma(\omega-1)]}%
{(2i+\omega-1)(2i+\omega+1)}
\\[0.2cm]
M_i&=&\dfrac{(i+1)(i+\omega-\alpha+1)(i+\omega-\beta+1)(i+\delta)}%
{(2i+\omega+1)(2i+\omega+2)}
\end{eqnarray*}
with $\omega=\gamma+\delta-1$.
The rescaled relaxation time $T/\tau$
is then the reciprocal of the lowest non-zero
solution of~\eqref{fraz_cont}.
In the case $d=2$ the conclusions are unchanged since 
the solution involving a logarithm in the neighbourhood of $z=0$
should be discarded.

We solved~\eqref{fraz_cont} numerically:
the continued fraction was computed by the modified Lentz's method
and the first non-zero solution was evaluated
by the root false position method \cite[see e.g.][]{Press}.

\section{Summary and discussion}
\label{sec:conclusions}

The goal of the paper was to investigate polymer stretching in a
turbulent flow within the context of a fully solvable model. The
statistical features of the Batchelor-Kraichnan flow allow us to
derive the complete form of the stationary p.d.f. of polymer
elongation for a general elastic force. When specializing to
finitely extensible polymers we recover the main properties of
polymer dynamics in real turbulent flows and compute the time of
relaxation to the stationary regime.

It should be noted that the velocity field we consider
is statistically isotropic. Together with the $\delta$-correlation in time,
this is a key assumption in order
to  derive a fully analytical solution of the problem.
In the experimental setup of \cite{Gerashchenko} the elastic turbulent flow is superimposed to a mean
shear flow. The long-time statistics of polymer extension in the presence of a mean shear
has been recently considered by \cite*{Chertkov et al.,CPT,Puliafito,Turitsyn}.

The main result of our study is the behaviour of the time of
relaxation to the steady state as a function of $\textit{Wi}$. At low~$\textit{Wi}$ the
transient relaxation time is an increasing function of $\textit{Wi}$, it is
maximum close to the coil--stretch transition, and eventually tends
to zero with increasing $\textit{Wi}$. Knowing the dependence of the transient time
on \textit{Wi} is relevant both for numerical simulations and experiments.
For example, in the former case,
the time required for uncorrelated polymer chains that are suddenly
exposed to the same flow to correlate is (implicitly) related to the
sharpness of stress gradients one can expect in the flow.  Hence, the
prediction of the transient time in our study is useful to
estimate the required grid spacing to fully resolve those gradients
(L. Collins 2005, private communication).
In the latter case,
the fact that the transient relaxation time is
especially long just below the coil--stretch transition implies
that within such range of~\textit{Wi} experimental measures are more
sensitive to statistical fluctuations.

Experiments
concerning the transient relaxation to the stationary regime can
investigate the time dependence of the conditional p.d.f.
$\mathcal{P}(R,t|R_0,0)$, which corresponds to the initial
condition peaked at the equilibrium size:
$\mathcal{P}(R,0|R_0,0)= \delta(R-R_0)$. Such initial condition
can be fixed experimentally as follows 
(A. Celani 2005, private communication). 
The p.d.f. of the extension
is constructed by following the motion of different polymer molecules
and collecting $R(t)$ for each molecule: one should then start
counting time only when the length of the corresponding polymer is
approximatively $R_0$. This is equivalent to selecting the initial state
where all molecules have the equilibrium extension, $R(0)=R_0$.

As for the transient relaxation time, this can be measured directly from the
time behaviour of the conditional moments of the extension:
$\overline{R^{n}}(t)=\int R^n\, \mathcal{P}(R,t|R_0,0)\, \mathrm{d}R$,
where $n$ is a positive integer. The conditional p.d.f. can indeed be
expanded as in~\eqref{eq:series} and the order of series and
integral can be interchanged in the definition of
$\overline{R^{n}}$ due to the integrability of 
$R^n p_k(R)$ and the uniform convergence of 
series~\eqref{eq:series} \cite[see e.g.][]{Smirnov}. 
Therefore, all the moments of the extension converge
to their stationary value with the same rate as the p.d.f. of the
extension.

To conclude, we believe that the results obtained
for the nonlinear dumbbell model are relevant for the
comprehension of polymer dynamics in turbulent flows at any
Weissenberg number. Moreover, we think that our study may stimulate
new experiments directed to investigate the transient relaxation to the
stationary regime.

\begin{acknowledgements}
We are deeply grateful to E. Bodenschatz, A. Celani, and L. Collins 
for very useful suggestions. We wish to thank
C.~M.~Ca\-scio\-la, M.~Chertkov, J.~Davoudi, E.~De
Angelis, B.~Eckhardt, A.~Mazzino, S.~Musacchio, J.~Schumacher, and
V.~Steinberg for illuminating discussions. This research has been
supported in part by the European Union under contract No.~HPRN-CT-2000-00162,
by the French-German program
PAI ``Procope''~07574NM, and by the Italian Ministry of Education,
University and Research under contract Cofin.~2003 (prot.~020302002038).
\end{acknowledgements}


\begin{thebibliography}{13}

\bibitem[Babcock \emph{et al.}(2003)]{Babcock}
\textsc{Babcock H. P., Teixeira R. E., Hur J. S., Shaqfeh E. S. \& Chu S.} 2003
Visualization of Molecular Fluctuations near the Critical Point
of the Coil--stretch Transition in Polymer Elongation,
{\it Macromolecules} {\bf 36}, 4544-4548.

\bibitem[Balkovsky \emph{et al.}(2000)Balkovsky, Fouxon \& Lebedev]{Balkovsky}
\textsc{Balkovsky E., Fouxon A. \& Lebedev V.} 2000
Turbulent Dynamics of Polymer Solutions. {\it Phys. Rev. Lett.}
{\bf 84}, 4765-4768.

\bibitem[Bird~\emph{et al.}(1977)]{Bird}
\textsc{Bird R. B., Hassager O., Armstrong R. C. \& Curtiss C. F.} 1977
{\it Dynamics of Polymeric Liquids, Kinetic Theory},
vol. II. Wiley.

\bibitem[Celani \emph{et al.}(2005)Celani, Musacchio \& Vincenzi]{Celani}
\textsc{Celani A., Musacchio S. \& Vincenzi D.} 2005
Polymer transport in random flow. {\it J. Statist. Phys.} {\bf 118}, 531-554.

\bibitem[Celani \emph{et al.}(2005)Celani, Puliafito \& Turitsyn]{CPT}
\textsc{Celani A., Puliafito A. \& Turitsyn K.} 2005
Polymers in linear shear flow: a numerical study. {\it Europhys. Lett.} 
{\bf 70}, 464-470.

\bibitem[Chertkov(2000)]{Chertkov}
\textsc{Chertkov M.} 2000 Polymer Stretching by Turbulence.
{\it Phys. Rev. Lett.} {\bf 84}, 4761-4764.

\bibitem[Chertkov \emph{et al.}(2005)]{Chertkov et al.}
\textsc{Chertkov M., Kolokolov I., Lebedev V. \& Turitsyn K.} 2005
Statistics of Polymer Extension in a Random Flow with Mean Shear.
{\em J. Fluid Mech.} {\bf 531}, 251-260.

\bibitem[De Gennes(1974)]{DeGennes}
\textsc{De Gennes P. G.} 1974 Coil--stretch transition of dilute flexible polymer
under ultra-high velocity gradients.
{\it J. Chem. Phys.} {\bf 60}, 5030-5042.

\bibitem[Erd\'elyi(1944)]{Erd}
\textsc{Erd\'elyi A.} 1944 Certain expansions of solutions of the Heun
equation. {\it Q. J. Maths (Oxford Series)} {\bf 15}, 62-69.


\bibitem[Falkovich \emph{et al.}(2001)Falkovich, Gaw\c{e}dzki \& Vergassola]{Falkovich}
\textsc{Falkovich G., Gaw\c{e}dzki K. \& Vergassola M.} 2001
Particles and fields in fluid turbulence.
{\it Rev. Mod. Phys.} {\bf 73}, 913-975.

\bibitem[Gerashchenko \emph{et al.}(2005)Gerashchenko, Chevallard \& Steinberg]{Gerashchenko}
\textsc{Gerashchenko S., Chevallard C. \& Steinberg V.} 2005
Single polymer dynamics: coil--stretch transition in a random flow.
{\it Europhys. Lett.} {\bf 71}, 221-227.

\bibitem[Gyr \& Bewersdorff(1995)]{Gyr}
\textsc{Gyr A. \& H. W. Bewersdorff H. W.} 1995
\emph{Drag reduction of turbulent flows by additives}.
Kluwer Academic.

\bibitem[Groisman \& Steinberg(2000)]{Groisman}
\textsc{Groisman A. \& Steinberg V.} 2000 Elastic turbulence in a polymer solution flow.
{\it Nature} {\bf 405}, 53-55.

\bibitem[Hur \emph{et al.}(2002)]{Hur}
\textsc{Hur J. S., Shaqfeh E. S., Babcock H. P. \& Chu S.} 2002
Dynamics and configurational fluctuations of single DNA
molecules in linear mixed flows. {\it Phys. Rev. E} {\bf 66},
011915.

\bibitem[Klyatskin \emph{et al.}(1996)Klyatskin, Woyczynski \& Gurarie]{Klyatskin}
\textsc{Klyatskin V. I., Woyczynski W. A. \& Gurarie D.} 1996
Short-time correlation approximations for diffusing tracers in random
velocity fields: A functional approach.
In {\it Stochastic Modeling in Physical Oceanography}, vol. 39 of
Progress in Probability,
R. J. Adler ed., pp.~221-270, Birkhauser.

\bibitem[Kraichnan(1968)]{Kraichnan_68}
\textsc{Kraichnan R. H.} 1968 Small-scale structure of a scalar field convected by
turbulence. {\it Phys. Fluids} {\bf 11}, 945-963.

\bibitem[Kraichnan(1974)]{Kraichnan_74}
\textsc{Kraichnan R. H.} 1974
Convection of a passive scalar by a quasi-uniform
random straining field.
{\it J. Fluid Mech.} {\bf 64}, 737-762.

\bibitem[Monin \& Yaglom(1975)]{Yaglom}
\textsc{Monin A. A. \& Yaglom A. M.} 1975 {\it Statistical Fluid Mechanics}.
MIT.

\bibitem[Perkins \emph{et al.}(1997)Perkins, Smith \& Chu]{Perkins}
\textsc{Perkins T. T., Smith D. E. \& Chu S.} 1997
Single Polymer Dynamics in an Elongational Flow.
{\it Science} {\bf 276}, 2016-2021.

\bibitem[Press \emph{et al.}(1993)]{Press}
\textsc{Press W. H., Flannery B. P., Teukolsky S. A. \&
Vetterling W. T.} 1993 \emph{Numerical Recipes}. Cambridge University.

\bibitem[Puliafito \& Turitsyn(2005)]{Puliafito}
\textsc{Puliafito A. \& Turitsyn K.} 2005 Single polymer
dynamics in shear flow. Accepted for publication in {\it Physica}~D.

\bibitem[Risken(1989)]{Risken}
\textsc{Risken H.} 1989
{\it The Fokker-Planck Equation, Methods of Solution and Applications}.
In Springer Series in Synergetics,  (H. Haken ed.)
Springer.

\bibitem[Ronveaux(1995)]{Ronveaux}
\textsc{Ronveaux A.} (ed.) 1995
{\it Heun's Differential Equations.} Oxford University.

\bibitem[Smith \emph{et al.}(1999)Smith, Babcock \& Chu]{Smith}
\textsc{Smith D. E., Babcock H. P. \&
Chu S.} 1999 Single-Polymer Dynamics in Steady Shear Flow.
{\it Science} {\bf 283}, 1724-1727.

\bibitem[Smirnov(1984)]{Smirnov}
\textsc{Smirnov V.} 1984
{\it Cours de Math\'ematiques Sup\'erieures}, Vol. IV, 2nd part. 
MIR. 

\bibitem[Thiffeault(2003)]{Thiffeault}
\textsc{Thiffeault J.-L.} 2003 Finite extension of polymers in turbulent flow.
{\it Phys. Lett.}~A {\bf 308}, 445-450.

\bibitem[Turitsyn(2005)]{Turitsyn}
\textsc{Turitsyn K. S.} 2005 Polymer dynamics in chaotic flows with strong shear component. Submitted to {\it Phys. Rev.}~E.

\bibitem[Whittaker \& Watson(1996)]{WW}
\textsc{Whittaker E. T. \& Watson G. N.} 1996
{\it A Course of Modern Analysis}, 4$^{\mbox{th}}$ ed.
Cambridge University.


\end{thebibliography}
\end{document}